\documentclass[conference]{IEEEtran}
\IEEEoverridecommandlockouts

\usepackage{cite}
\usepackage{amsmath,amssymb,amsfonts}
\usepackage{graphicx}
\usepackage{textcomp}
\usepackage{xcolor}
\usepackage{multirow}
\usepackage{hyperref}
\usepackage[ruled,vlined]{algorithm2e}
\SetKwSty{bfseries}{}
\SetArgSty{textnormal}

\usepackage{afterpage}
\def\BibTeX{{\rm B\kern-.05em{\sc i\kern-.025em b}\kern-.08em
    T\kern-.1667em\lower.7ex\hbox{E}\kern-.125emX}}

\begin{document}

\title{EdgeMM: \underline{M}ulti-Core CPU with Heterogeneous AI-Extension and Activation-aware Weight Pruning for \underline{M}ultimodal LLMs at Edge
%\title{EdgeMM: Heterogeneous AI-Extended \underline{M}ulti-Core CPU with Bandwidth Optimization and Allocation for \underline{M}ultimodal LLMs at Edge
\thanks{*Corresponding author.}
}
%Activation-Aware
\author{
\IEEEauthorblockN{Kangbo Bai}
\IEEEauthorblockA{\textit{School of Integrated Circuits} \\
\textit{Peking University}\\
Beijing, China \\
baikangbo@pku.edu.cn}

\and
\IEEEauthorblockN{Le Ye}
\IEEEauthorblockA{\textit{School of Integrated Circuits} \\
\textit{Peking University}\\
Beijing, China \\
yele@pku.edu.cn}

\and
\IEEEauthorblockN{Ru Huang}
\IEEEauthorblockA{\textit{School of Integrated Circuits} \\
\textit{Peking University} \\
Beijing China \\
ruhuang@pku.edu.cn}

\and
\IEEEauthorblockN{Tianyu Jia\textsuperscript{*}}
\IEEEauthorblockA{\textit{School of Integrated Circuits} \\
\textit{Peking University}\\
Beijing, China \\
tianyuj@pku.edu.cn}
%\vspace{-50pt}
}
% \author{Kangbo Bai\textsuperscript{1}, Yanchi Dong\textsuperscript{1}, Le Ye\textsuperscript{1,2}, Ru Huang{1}, Tianyu Jia\textsuperscript{1,*}\\
% \textsuperscript{1} School of Integrated Circuits, Peking University, Beijing, China \\  \textsuperscript{2} Advanced Institute of Information Technology of Peking University, Hangzhou, China \\
%   *Corresponding\ Author:\ tianyuj@pku.edu.cn
% \vspace{-10pt}
%City, Country email address or ORCID} and
%City, Country email address or ORCID} and
%\textit{name of organization (of Aff.)} City, Country email address or ORCID}
%}

\maketitle

\begin{abstract}
%The integration of AI instruction extension has emerged as a notable trend to boost the AI processing capabilities of CPUs. Enhancing processing parallelism through many-core architectures is effective in dealing with the rapid expansion of large language models (LLMs). In batched LLMs inference, general matrix multiplication (GEMM) and matrix-vector multiplication (GEMV) are bounded by computation and memory access respectively. To address this discrepancy, we construct heterogeneous coprocessors with diverse compute-to-memory ratios, based on systolic arrays and SRAM Computing-In-Memory (CIM) macros respectively.  Effective data management and scheduling are also proposed to optimize data transfer and hardware utilization.
Emerging multimodal LLMs (MLLMs) exhibit strong cross-modality perception and reasoning capabilities and hold great potential for various applications at edge. However, MLLMs typically consist of a compute-intensive modality encoder and a memory-bound LLM decoder, leading to distinct bottlenecks for hardware designs. In this work, we present a multi-core CPU solution with heterogeneous AI extensions, which are based on either the compute-centric systolic array or memory-centric digital compute-in-memory (CIM) co-processors. In addition, dynamic activation-aware weight pruning and bandwidth management are developed to enhance bandwidth efficiency and core utilization, improving overall performance. We implemented our solution using commercial 22nm technology. For representative MLLMs, our evaluations show EdgeMM can achieve 2.84$\times$ performance speedup compared to laptop 3060 GPU.
%reaching 138 tokens/s throughput and 0.28 token/J efficiency. 
\end{abstract}

\section{Introduction}
\begin{figure}[t]
  \centering
  \includegraphics[width=\linewidth]{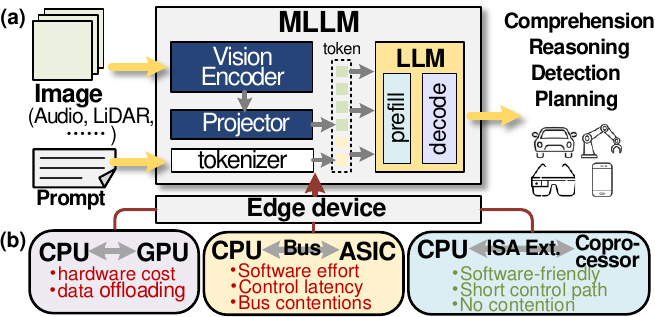}
  \vspace{-10 pt}
  \caption{MLLMs and hardware architecture design choices.}
  \label{fig:Comparison}
  \vspace{-10 pt}
\end{figure}

Multimodal large language models (MLLMs) exhibit strong perception, reasoning, and planning abilities, with promising applications in autonomous driving \cite{AD}, embodied AI \cite{embodied}, mobile \cite{chu2023mobilevlm}, and AR/VR. As shown in Fig.~\ref{fig:Comparison} (a), MLLMs take in various modalities of input, such as image, audio, and LiDAR, which are processed by modality encoder %e.g. CLIP \cite{clip}, 
to generate token representations. Following LLM and post-processing are performed for  
%vision, auditory, and radar sensors through encoders, and perform inference via a language model (LM) guided by text prompts, to complete 
tasks such as comprehension, object segmentation \cite{sam4mllm}, or trajectory planning \cite{lmdrive}. 
Due to the application demands for privacy and real-time responses, deploying MLLMs on edge devices is highly desired. 

%Nevertheless, MLLM inference on edge devices is highly challenging due to constraints on hardware resources, latency, and power consumption.
Achieving low-latency inference of MLLMs at edge is challenging. 
First, the hardware resource is quite constrained at edge. 
%For example, a device using a smartwatch battery or a button cell battery can only maintain 60mA for two hours \cite{yele_aiot}.
Second, different phases in MLLMs exhibit opposite characteristics and performance bottlenecks. %\cite{kim2023full}.
%Our analysis for MLLMs also reveals that the Decoder has similar parameter count with the encoder, while the FLOPs of the latter is two orders of magnitude higher than the former.
For instance, the encoder network is compute-intensive, while the LLM-decoding phase is bounded by the memory bandwidth.
Such workload behaviors limit performance improvements of certain accelerators tailored for particular Transformers \cite{nv_acc,edgebert}. %\cite{RASA,modeling,VEGETA,xpulpNN}. 
%A single hardware configuration cannot meet diverse demands. 
Meanwhile, the prior optimization techniques for long-context LLMs are not suitable for edge MLLMs, which process short token sequences with streaming inputs.
%especially in real-time interactive applications with streaming inputs. 
%For example, most prior work \cite{RASA,modeling,VEGETA,xpulpNN} only focused on a single core with coprocessor, which fails to meet practical demands: %due to the following shortcomings. %The inference of LLMs requires massive parallelism and has its unique challenges. 
%First, a single core ca struggles to achieve LLM inference within the required latency for its restricted performance and parallel processing capabilities.%The parallelism and performance of single-core are limited, making it difficult to complete MLLM inference within the desired latency. 
%Second, homogeneous compute core cannot accommodate diverse computational and memory requirements.%the arithmetic intensity of different LLMs stages, i.e. prefill and decoding, requires quite diverse hardware resources for computation and memory bandwidth. Increasing computational resources blindly does not alleviate memory bottlenecks and can adversely affect energy efficiency.
%and homogeneous hardware configurations cannot adapt diverse computation and bandwidth requirements.%Second, to cope with the varying bandwidth and computing characteristics at different stages of transformer inference, heterogeneous architecture is an effective strategy for PPA optimization\cite{attacc,neupim,patel2024splitwise}. 

There are multiple hardware architecture choices for edge devices, as shown in Fig.~\ref{fig:Comparison}(b). 
CPU remains the most common hardware in edge devices, while it has limited parallelism for AI tasks.
%CPU is the most primary component in edge devices. However, it is clear that the existing CPU architecture lacks parallelism and cannot meet the surging demands for AI workloads. 
The use of GPU on edge devices is severely limited by cost, and data offloading from CPU to GPU may become a system bottleneck \cite{HeteGen}. ASIC accelerator is also a popular solution \cite{ASIC}, which is connected to host CPU through memory-mapped interface.
%, allowing the host core controls accelerators by specific memory accesses. 
However, ASIC accelerators require corresponding compiler development, adding huge engineering effort for software developers. They also introduce bus access with increased control latency and bus contention \cite{blueface}.
%, e.g. Transformer-based large language models (LLMs) \cite{transformer}. 
%GPUs and ASIC accelerators utilize massive parallel architecture to enhance AI performance, while offloading substantial data from CPU to GPU may become a bottleneck for overall performance \cite{HeteGen}.

In recent years, AI-target ISA extension with in-CPU coprocessors has emerged as a promising solution for AI tasks. 
%In recent years, CPU-Coprocessor is a promising solution for AI workloads by developing in-CPU coprocessor and matrix ISA extension.
%instruction set architecture (ISA) extensions. 
For example, Intel introduced its advanced matrix extensions (AMX) \cite{intel-amx-web} in the Xeon processors \cite{Xeon, intel2024isca}. 
Arm developed scalable matrix extension (SME) in Armv9-A \cite{arm-sme}. 
T-head proposed open-source RISC-V matrix extension \cite{t-head-mme}. 
%To support the AI-specific ISA extensions, in-CPU coprocessors are needed.
%Fig.~\ref{fig:Comparison} shows detailed comparisons of different architectural paths. %Compared to conventional CPUs or GPUs, 
%To support these instruction extensions, it is necessary to incorporate coprocessors into the CPUs. 
%As shown in fig~\ref{fig:Comparison}(b), typically, 
%As shown in Fig.~\ref{fig:Comparison}, typical ASIC accelerators are connected to CPU host cores through memory-mapped (MM) interfaces, allowing the host cores control accelerators by specific memory accesses. However, MM interface introduces bus access, increasing the latency of control and response, and may cause bus resource contentions \cite{blueface}. 
At architecture level, the host CPU cores decode and dispatch these extended instructions to coprocessors for execution via direct-linked interfaces.
%, enabling the host core to dispatch extended instructions to coprocessors for execution. %For AI extenstion in CPUs, the host cores and AI coprocessors are connected through direct-linked core-accelerator (DL-CX) interfaces, by which the host core can directly dispatch extended instructions to coprocessors for execution. 
%Coprocessors have following advantages 
Compared to conventional bus-attached accelerators, coprocessors have a shorter control path and avoid bus contention. 
Moreover, this architecture can be merged into existing CPU ISA and inherit the compiler stack \cite{intel2024isca}.
%, providing a friendly ecosystem for development . 
%Most previous works \cite{gemmini-dac,RASA, VEGETA,xpulpNN} in academia focus solely on single-core homogeneous coprocessor, lacking of system-level architecture exploration for more complex tasks, such as MLLMs.

%Under the constraints on hardware resources, latency, and power consumption, MLLM inference on edge devices is highly challenging

%To address the above challenges, 
In this work, we explore a multi-core CPU solution for edge MLLMs with heterogeneous AI-extension.
%and an activation-aware weight pruning bandwidth optimization technique. 
%Based on the characteristic of MLLMs, we develop two types of AI-extension architectures, i.e. digital systolic array and compute-in-memory (CIM), in coprocessors for MLLMs. 
For compute-intensive general matrix multiplication (GEMM),
%in modality encoder and LLM-prefill phase, 
systolic array is utilized as the coprocessor. 
Alongside, for memory-bound matrix-vector multiplication (GEMV), which dominates LLM-decoding phase, 
%which is unsuitable for systolic data flow, 
area- and power-efficient digital compute-in-memory (CIM) macros are adopted. 
These coprocessors have different dataflows and resource distributions, coupled with RISC-V ISA and programming model for easy deployment.
%incorporating computation units closely coupled with memory and abundant on-chip memory to support efficient data transfer.
%to accelerate LLM short-context inference which is overlooked previously but important at edge application, 
Furthermore, a dynamic activation-aware weight pruning is developed along with hardware support to alleviate memory bandwidth pressure. 
%To enhance data transmission, we practically analyze the characteristics of the direct memory access (DMA) module and propose targeted measures to alleviate bandwidth pressure by increasing CIM data memory capacity and optimizing data management.%to avoid array utilization decrease that occurs with systolic arrays, and save area for larger data memory to alleviate bandwidth pressure. 
At the system level, 
%heterogeneous cores introduce new challenges in scheduling. 
the DRAM bandwidth can be dynamically managed to fully utilize heterogeneous cores for diverse sequence lengths.
%To fully harness the heterogeneous cores across diverse scenarios, a dynamic bandwidth allocation and scheduling scheme has also been developed. %To further accelerate the memory-bound decode phase in LM, we also introduce the bandwidth distribution and activation-aware weight pruning.

The contributions of this paper are summarized as follows: 
\begin{itemize}
\item We analyze the attributes of edge MLLMs, revealing that different phases encounter distinct bottlenecks in computation and memory access.
\item We develop the heterogeneous extension for multi-core CPU, employing the systolic array and CIM macro tailored for GEMM and GEMV.
%To address the diverse bottlenecks caused by computation and memory access, systolic array-based and CIM-based heterogeneous coprocessors are integrated in multi-core CPU for GEMM and GEMV.%We introduce a heterogeneous multi-core CPU architecture with AI extensions for edge MLLMs. To address computation and memory access bottlenecks, systolic array-based and CIM-based coprocessors are designed for GEMM and GEMV. 
\item We present an activation-aware weight pruning and bandwidth management, leveraging the limited bandwidth and heterogeneous cores to accelerate edge MLLM. %To leverage limited bandwidth for edge MLLM, activation-aware dynamic weight pruning is proposed with hardware support.
%\item To address challenges of scheduling heterogeneous cores in edge MLLM, a bandwidth and workload management is utilized for better hardware utilization.
\item EdgeMM is implemented and evaluated at 22nm, achieving a 0.217 token/J efficiency and 138 tokens/s throughput, which is 2.84$\times$ better than RTX 3060 GPU.
\end{itemize}

\begin{table}[t]
      \caption{Representative MLLMs and efficient edge MLLMs.}
    \label{mllm_table}
    \vspace{-5 pt}
  \resizebox{\linewidth}{!} {
\begin{tabular}{llll}
\hline
MLLMs       & Visual Encoder                                                                         & Projector & Language Model      \\ \hline

Emu2-Chat \cite{Emu}     & EVA \cite{EVA}                                                                                    & MLP        & LLaMa (33B)               \\
LLaVA \cite{llava}      & CLIP ViT/L-14 \cite{clip} (0.3B)                                                                   & MLP       & Vicuna \cite{chiang2023vicuna} (7B, 13B)    \\ \hline
MobileVLM \cite{chu2024mobilevlm}   & CLIP ViT/L-14 \cite{clip} (0.3B)                                                                   & LDP       & MobileLLaMA \cite{chu2024mobilevlm} (2.7B)  \\
TinyGPT-V \cite{yuan2023tinygpt}   & EVA                                                                                    & Q-former \cite{q-former} & Phi-2 \cite{javaheripi2023phi} (2.7B)        \\
SPHINX-Tiny \cite{gao2024sphinx}& \begin{tabular}[c]{@{}l@{}}CLIP ViT/L-14 \cite{clip}, CLIP ConvNeXt \cite{clip}\\  + DINOv2(0.4B) \cite{oquab2023dinov2}\end{tabular} & MLP       & TinyLlama \cite{zhang2024tinyllama} (1.1B)    \\
DeepSeek-VL \cite{lu2024deepseek} & SigLIP-L \cite{zhai2023sigmoid} (0.4B)                                                                        & MLP       & DeepSeek-LLM \cite{lu2024deepseek} (1.3B) \\
Karmavlm \cite{karmavlm}   & \begin{tabular}[c]{@{}l@{}}SigLIP-so \cite{zhai2023sigmoid} (0.4B), \\ CLIP ViT/L-14 \cite{clip} (0.3B)\end{tabular}      & MLP       & Qwen1.5 \cite{bai2023qwen} (0.5B)      \\ \hline
\end{tabular}
}
\vspace{-10pt}
\end{table}

\section{Background and Motivations}
\subsection{Multimodal Large Language Models}
MLLM builds on standard LLM by incorporating multimodal data for input and output.
%Unlike conventional multimodal models, MLLMs relys on LLM with billionscale parameters and exhibits remarkable capabilities.
In recent two years, leading MLLMs such as Gemini \cite{team2023gemini}, GPT-V \cite{achiam2023gpt}, LLaVA \cite{llava}, InstructBLIP \cite{instructBLIP} have emerged, impacting both industry and academia. Combining LLMs' reasoning with multimodal perception, they exhibit remarkable performance.

Fig.~\ref{fig:Comparison}(a) illustrates the typical backbone architecture of MLLM, %using vision modality as an example. 
which is composed of modality encoder, projector, and LLM. Take visual modality as an example. The features are extracted from input images by Transformer-based vision encoder, e.g. CLIP \cite{clip}, %SigLIP \cite{zhai2023sigmoid}, 
DINOv2 \cite{oquab2023dinov2} and then aligned with language model via a projector, e.g. MLP, Q-former \cite{q-former}, to generate vision tokens. 
The vision tokens and text tokens from the prompt are jointly fed into the LLM. The inference of LLM consists of two phases, i.e. prefill and decoding. %\cite{kim2023full}. 
In the prefill phase, attention is applied between all tokens, generating the first output token and key-value (KV) cache. In the decoding phase, only the last token is taken as input, autoregressively generating the next token and updating the KV cache. 
The generated tokens can be converted into language as output or further processed by downstream tasks, e.g. AD \cite{AD}.

Tab.~\ref{mllm_table} lists a few representative open-source MLLMs and their model architectures. 
MLLMs \cite{Emu,llava} often adopt LLMs with parameter amounts surpassing 7B.
%The upper section of Table ~\ref{mllm_table} presents mainstream MLLMs, yet their parameter amounts surpass 7B, hindering edge implementation. 
Meanwhile, there is also a model development trend for lightweight MLLMs to assist the deployment at edge. %\cite{chu2024mobilevlm, llava-phi, zhou2024tinyllava, yuan2023tinygpt, gao2024sphinx, karmavlm, lu2024deepseek}.
%which are listed in the lower part of Table ~\ref{mllm_table}.
%The lower section enumerates the latest efficient MLLMs , 
These edge MLLMs adopt compressed LLMs with less than 3B parameters, such as Phi2-2.7B \cite{javaheripi2023phi}, %MobileLlaMa-2.7B \cite{chu2024mobilevlm}, 
TinyLlama-1.1B \cite{zhang2024tinyllama}, Qwen1.5-0.5B \cite{bai2023qwen}. %significantly reducing the parameter amounts and being more friendly for edge inference deployment. 
It is also worth pointing out that these edge MLLMs also demonstrate comparable performance compared to large-scale MLLMs in many benchmarks \cite{VQA, SQA, SEED, MMB}. %For instance, Bunny-v1.1-4B \cite{bunny} achieve similer or even higher score than GPT4-V \cite{achiam2023gpt} and LLaVA \cite{llava} in %VQA \cite{VQA}, SQA \cite{SQA}, SEED \cite{SEED} or MMB \cite{MMB} benchmarks.

%In edge scenarios, MLLMs usually have short input and output token sequences, but they face strict latency requirements for inference especially in real-time interactive applications like AR/VR and autonomous driving. Therefore, conventional long-context decoding focused LLMs inference optimizations are not suitable for edge MLLMs. 
\subsection{Model Profiling and Analysis}
\label{sec:profiling}
%\subsection{Profiling of MLLMs}
\begin{figure}[t]
  \centering
  \includegraphics[width=\linewidth]{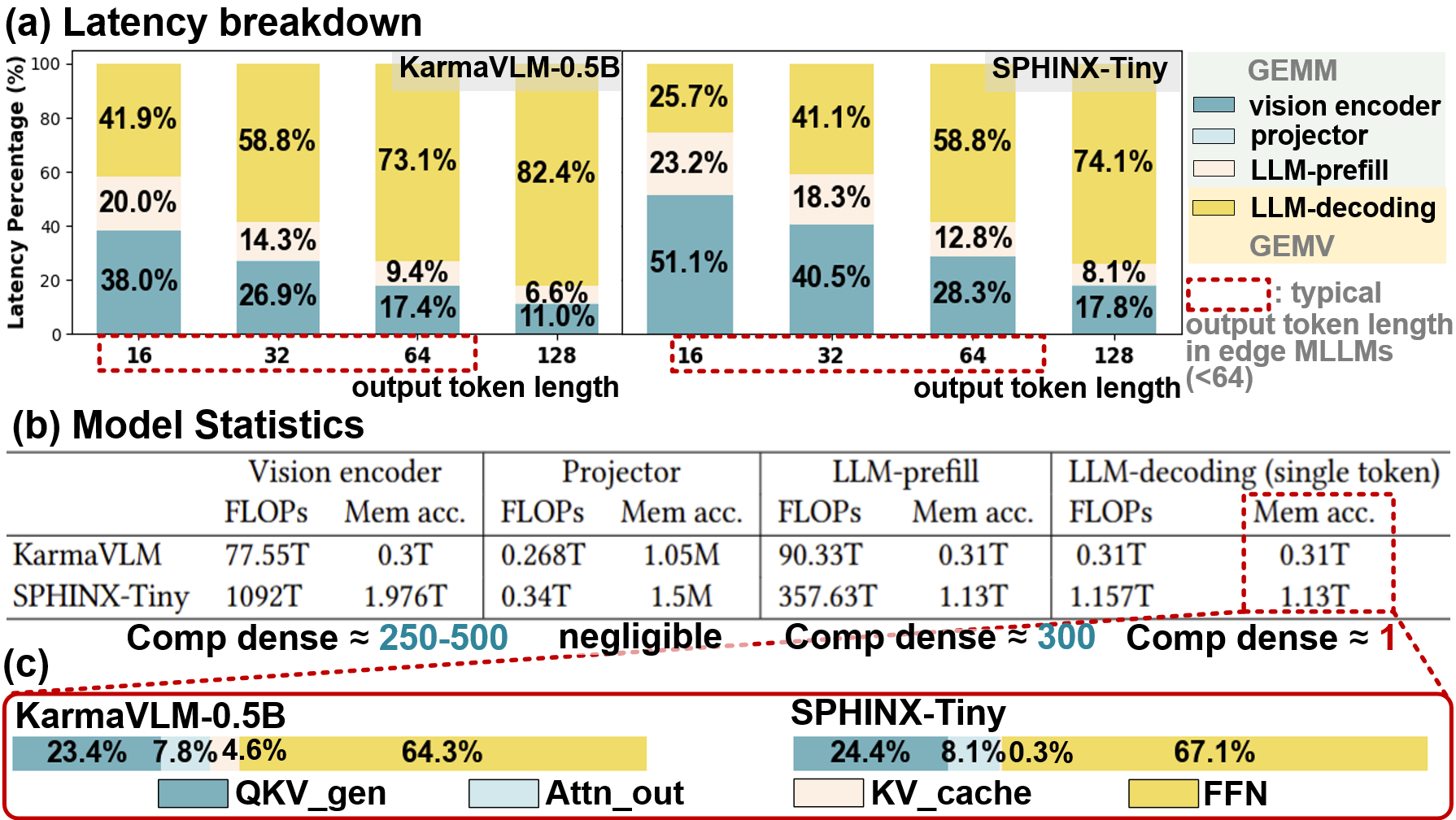}
  \vspace{-20 pt}
  \caption{Workload analysis of two MLLMs: (a) latency breakdown, (b) model statistics, (c) memory accesses.}
  \label{fig:workload}
  \vspace{-15 pt}
\end{figure}

To better understand the workload characteristics, we conduct profiling on two representative MLLMs: SPHINX-Tiny \cite{gao2024sphinx} and Karmavlm \cite{karmavlm}. 
The input token length is about 300, primarily made up of vision tokens.
%and text tokens are relatively few. 
%To understand the model characteristics of MLLMs, we perform the profiling analysis for typical edge MLLMs, including SPHINX-Tiny \cite{gao2024sphinx} and Karmavlm \cite{karmavlm}.
The inference latency breakdown on Nvidia 3060 GPU with varying output token lengths is shown in Fig.~\ref{fig:workload}(a). 
%It is clearly observed that more LLM decoding iterations are executed and result in a larger portion in overall latency with increased output token length. 
An increase in output tokens leads to more LLM-decoding iterations, enlarging the proportion of this phase in overall latency. Alongside, the vision encoder and LLM-prefill also contribute to the overall latency notably in edge applications with short token lengths. The latency of the projector is negligible.
%the workloads of the vision encoder and LLM-prefill stay the same 

 %It is observed that the vision encoder, LLM prefill and decoding share a notable portion of the inference latency. 
%As the profiled parameter and operation numbers in Fig~\ref{fig:workload} (b), each phase differs significantly in the computing and memory characteristics. 
Fig.~\ref{fig:workload}(b) also indicates the computing and memory attributes differ significantly during different phases. In vision encoder and LLM-prefill, multiple token embedding vectors share weight matrices forming computation-intensive GEMM. 
Meanwhile, LLM-decoding performs the operations between the entire weight matrix and a single embedding vector, resulting in low-computational-density GEMV. %Hence the computation of MLLMs can be divided into two types: GEMM-based vision encoder, projector, LLM prefill, and GEMV-based LLM decoding.
%Table ~\ref{MLLM_profile} list the profile results.  
%as many MLLMs' vision encoders generate this count of vision tokens, 
% For LLM-decoding, we present static result for generating a single token. 
%The overall amount for a output sequence scales linearly with the output token length. 
%The vision encoder and LLM-prefill exhibit comparable computational density, both in the hundreds in our tests.
The LLM-decoding employs the same model parameters as LLM-prefill but involves two orders of magnitude fewer FLOPs, revealing an obvious memory-bound bottleneck. As shown in Fig.~\ref{fig:workload}(c), weight matrices dominate the memory access, with FFN consuming the largest portion as the channel dimension is typically several times larger than the model dimension \cite{kim2023full}. 
The KV cache only contributes a small portion of memory access due to the short token length in edge MLLMs.
%However, the parameter count for a single decoding operation is similar to that of the LLM-prefill as shown in Fig~\ref{fig:workload} (c). During overall autoregressive sequence generation, the FLOPs of decoding still not exceeding that of the GEMM-based phases. Nevertheless, the number of DRAM memory accesses during decoding is significantly higher, revealing a clear memory-bound performance bottleneck.

%\subsection{Activation Vector Sparsity Analysis}

\begin{figure}[t]
   \centering
   \includegraphics[width=\linewidth]{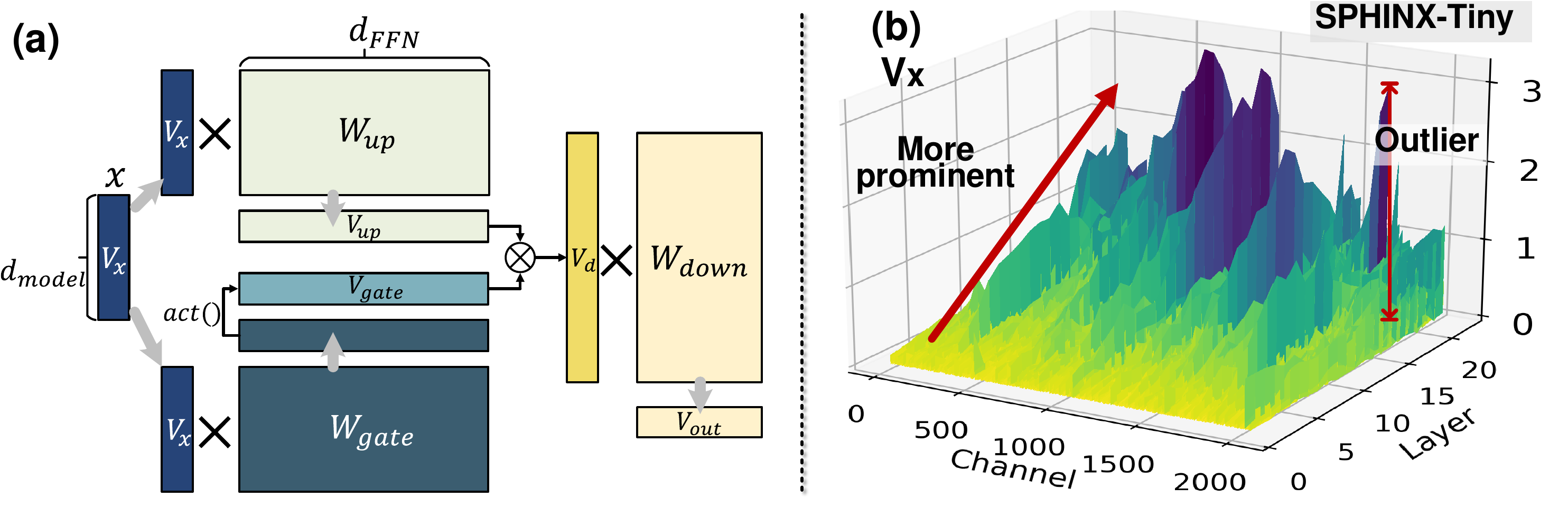}
   \vspace{-15pt}
   \caption{Gated-MLP and the activation vector sparsity in FFN.}
   \label{fig:activation}
   \vspace{-15pt}
 \end{figure}

%Leveraging sparsity can effectively optimize memory access and computation in LLM. As presented in Section 3.1, LLM-decoding is significantly constrained by bandwidth, and 
Since the majority of DRAM access is occupied by FFN weight matrices, we further analyze the characteristics of this portion, e.g. sparsity.
Currently the mainstream LLMs, e.g. Llama series \cite{touvron2023llama}, Mistral \cite{jiang2023mistral}, Qwen \cite{bai2023qwen}, use the gated-MLP as FFN. Fig~\ref{fig:activation} (a) shows the detailed process, 
which contains three GEMV, and the calculation formula is as follows:
\begin{equation}
   FFN(V_{x})= ((V_{x}*W_{up} )\circ (act(V_{x}*W_{gate})))*W_{down} 
\end{equation}
where $V_{x}$ is the input vector, $W_{up} ,W_{gate}$ are $d_{FFN}\times d_{model}$ weight matrices. $Act()$ is the activation function. %$W_{up} ,W_{gate} \in R^{d_{model}\times d_{FFN}}$, $W_{down} \in R^{d_{FFN}\times d_{model}}$. $act()$ is the activation function. 
%The input is a $d_{model}$-dimension vector $x$, which is multiplied by two $d_{model} \times d_{FFN}$ weight matrices $W_{up}$ and $W_{gate}$. The resulting vector v1 is processed by the activation function and then multiplied with v2 element-wise. The generated v3 is multiplied with $W_{down}$ to get the output vector.
In fact, the activation vectors ($V_{x}$, $V_{d}$) in GEMV exhibit notable sparsity across channels. Fig.~\ref{fig:activation}(b) shows profiled magnitudes of $V_{x}$ %($V_{x}$, $V_{up}$, $V_{gate}$, $V_{d}$) 
during a token generation in SPHINX-Tiny \cite{gao2024sphinx}. 
The x, y, and z axes denote the decoder layer index, vector channels, and absolute magnitudes. As presented, these activation vectors exhibit notable sparsity for most channels, with a few outlier channels that can be easily masked out. %which is also observed by previous work \cite{LLM.int8()}. 
Moreover, as the layer index increases, these outliers become more prominent. %These characteristics are leveraged by proposed pruning method which is detailed in Section 5.1.
%Based on the above insights, we develop a multi-core CPU design with heterogeneous in-CPU acceleration architectures. 
%To further enhance the memory-bounded GEMV operations, an activation-aware weight pruning method is implemented. %in the GEMV acceleration.
%To better support balanced execution of heterogeneous cores, the dynamic bandwidth allocation is also supported for scheduling.
%\textcolor{red}{Based on xxx observations/considerations, this motivates us xxx}
%To deploy MLLMs on edge devices, many recent works incorporate compressed language model, significantly reducing the number of model parameters to while maintaining excellent performance.
%Currently, most prevalent LLMs are based on decoder-only Transformer architectures. The inference of these models contains two phases: Prefill and Decoding, as illustrated in Fig~\ref{fig:transformer}. In Prefill phase, the tokens of input query are processed together to generate the first output token, which mainly performs GEMM calculations. In Decoding phase, the last token passes through multiple decoder blocks to produce a new token, and this process is repeated until the full output sequence is finalized. Decoding phase is consisted of memory-bound GEMV calculations and dominates the majority of over all inference time.

The above analysis guides us to rethink the optimal architecture for edge MLLMs with the following design insights.
\textbf{Insight 1: Heterogeneous coprocessors for GEMM and GEMV are desired.} 
%It is clear that different acceleration methods are desired for these distinct GEMM and GEMV phases. 
The performance of compute-bound GEMM can be enhanced by increasing computing resources, while memory-bound GEMV requires addressing bandwidth.
%Prior optimizations are unsuitable for the decoding \cite{dao2022flashattention}, or only reduce computation \cite{wang2021spatten, qin2023fact}, or need weight modification \cite{li2023losparse, qin2024mecla}. 
\textbf{Insight 2: Activation-aware weights pruning is promising.} 
%are unsuitable for the decoding \cite{dao2022flashattention, wang2021spatten}, or 
Many prior work focuses on long-context KV cache optimization \cite{Zhang2023H2OHO}. For short-context edge MLLMs, the weight matrices of FFN are key constraints, while the sparsity in most activation channels offers great potential for optimization. %while KV cache only contribute little portion of memory usage for edge MLLMs with relative short token length.
%and compression method, e.g. GQA \cite{ainslie2023gqa}. 
%Transferring such a large matrix solely for a GEMV is costly, thus improving this is key to boosting performance. 
\textbf{Insight 3: Scheduling for output token lengths is needed.} Since the workload proportion varies with the output token length, dynamic scheduling is also strongly desired to maintain the balanced workload latency of heterogeneous cores.

\section{EdgeMM Architecture Design}
\subsection{Architecture Overview}

\begin{figure}[t]
  \centering
  \includegraphics[width=\linewidth]{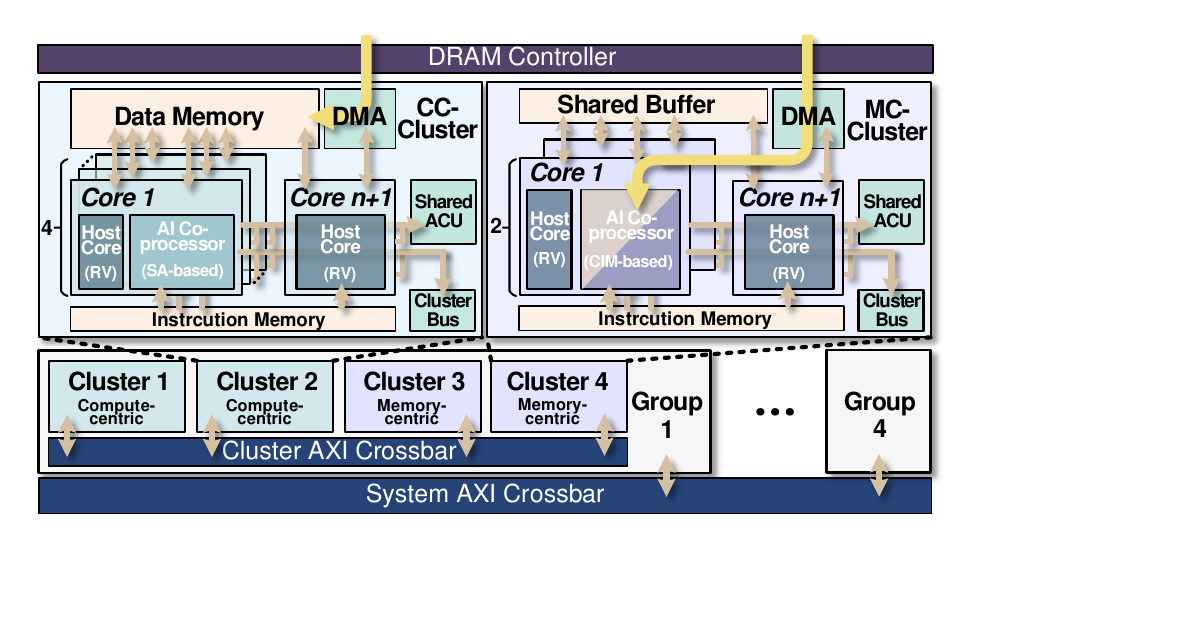}
  \vspace{-15pt}
  \caption{Architecture overview of EdgeMM design.}
  \label{fig:overview}
  \vspace{-15pt}
\end{figure}

To accommodate the mentioned characteristics of edge MLLMs, 
%mentioned in Section 2 and address diverse performance bottlenecks, 
we present a multi-core CPU architecture, i.e. EdgeMM, with heterogeneous AI extensions, 
%Multiple CPU cores with AI extensions are grouped as either compute-centric (CC) or memory-centric (MC) CPU clusters to process compute-bound GEMM and memory-bound GEMV operations, respectively. 
as shown in Fig.~\ref{fig:overview}. 
EdgeMM adopts a hierarchical multi-core architecture based on
%has emerged as a new trend \cite{zaruba2020snitch, mempool, jung2024scalable}, we adopt this architecture based on 
the open-source Snitch\_cluster \cite{zaruba2020snitch}. %for the MLLM deployment at edge. %It consists of multi-core RISC-V CPUs, with two heterogeneous types of extensions, i.e. \textit{i} compute-central clusters and \textit{j} memory-central clusters.
Two types of accelerator architecture %which are based on either systolic array or digital CIM macro, 
are incorporated as coprocessors for compute-centric (CC) cores and memory-centric (MC) cores.
%for the compute-bound phases and memory-bound phases in MLLMs. %within the CPU cores. 
Specifically, CC-cores utilize compute-dense systolic array (SA)-based coprocessors for GEMM. While MC-cores utilize memory-abundant digital CIM, which integrates the compute units within SRAM banks, as coprocessors for GEMV.

%The cores in CC-cluster employ systolic array (SA)-based AI extension, while the cores in MC-cluster utilize CIM-based architecture. 
%The design details of these two heterogeneous extension architectures will be explained the following section. 
For CC-cores, cores in a cluster share the instruction memory and data memory. In contrast, the data memory and compute array in MC-cores are integrated as a CIM macro in each core, and a small shared buffer is equipped for inter-core data transfer. 
Every cluster has a distributed DMA module, all of which are linked to the DRAM controller for data transfer to on-chip data memory. %The DMA module is responsible for transmitting data from the off-chip DRAM to the data memory. %The AXI buses connect each clusters and groups in a hierarchical way. 
Auxiliary compute units (ACU), e.g. 32-bit multiplier and divider, are also shared among cores in a cluster for uncommon calculations.

The entire chip contains 4 groups, each consisting of 2 CC-clusters and 2 MC-clusters.  % the compute-central cluster(CC-cluster) and the memory-central cluster(MC-cluster), with counts of \textit{i} and \textit{j} respectively. 
%Clusters are connected by the system AXI bus.
%The acceleration clusters share identical coprocessor interfaces, while the compute-central clusters employ systolic array-based coprocessors and memory-central clusters utilize CIM-based coprocessors. 
Each CC-cluster and MC-cluster contains 4 CC-cores and 2 MC-cores respectively. These cores integrate an area-efficient RISC-V host core \cite{zaruba2020snitch} for control and an extended coprocessor for AI computing. 
Each cluster also has a dedicated host core to control the DMA module. 
The groups and clusters are connected by the hierarchical AXI crossbars.
The hardware architecture can also be scaled by changing architecture parameters.
%These cores  \textit{n} equals 8 and for MC-cluster \textit{n} equals 4.
%The AXI bus connect each clusters and each groups in a hierarchical manner.
%The first \textit{n} cores in each cluster contain the RISC-V based host and AI extensions for computation acceleration, 
%named as compute core, 
%while the last core adopts a tiny core \cite{zaruba2020snitch} and only works as the host for control and DMA.

%Since host cores are only responsible for controlling and sending requests to coprocessors, snitch \cite{zaruba2020snitch}, a tiny processor with high power and area efficiency is chosen. 
 %to have flexible architecture evaluation and RTL generation. 
%A cluster consists of \textit{n}+1 cores, for CC-cluster \textit{n} = 8 and for MC-cluster \textit{n} = 4. %can be different for two kinds of clusters. 
 %In a compute core, the host core is only responsible for controlling and sending requests to the coprocessor, and the coprocessor handles the actual calculation. 

\subsection{CPU Extension for AI Acceleration}

\begin{figure}[t]
  \centering
  \includegraphics[width=\linewidth]{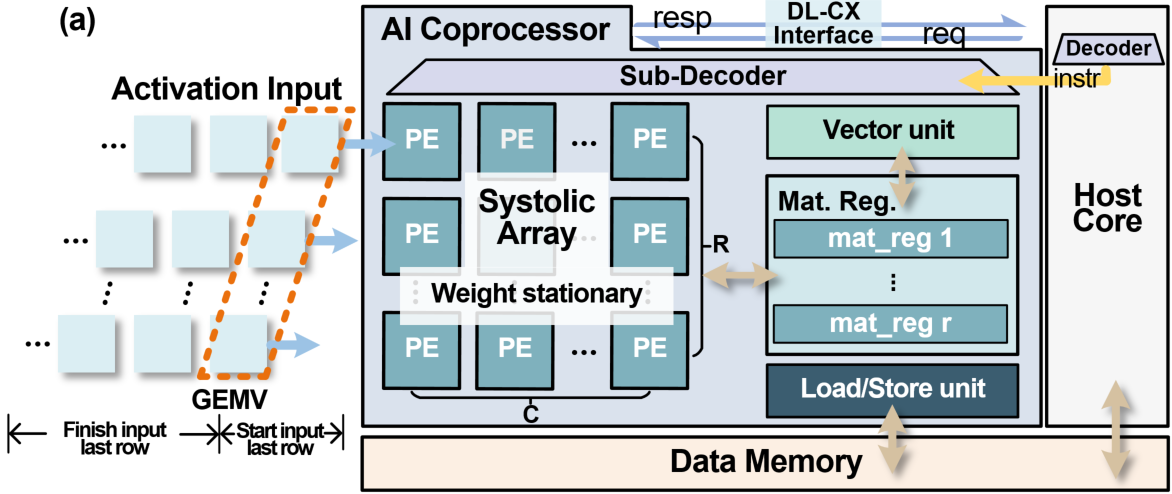}
  \vspace{-15 pt}
  \caption{Compute-centric core with Systolic Array.}
  \label{fig:systolic}
  \vspace{-10 pt}
\end{figure}

\begin{figure}[t]
  \centering
  \includegraphics[width=0.98\linewidth]{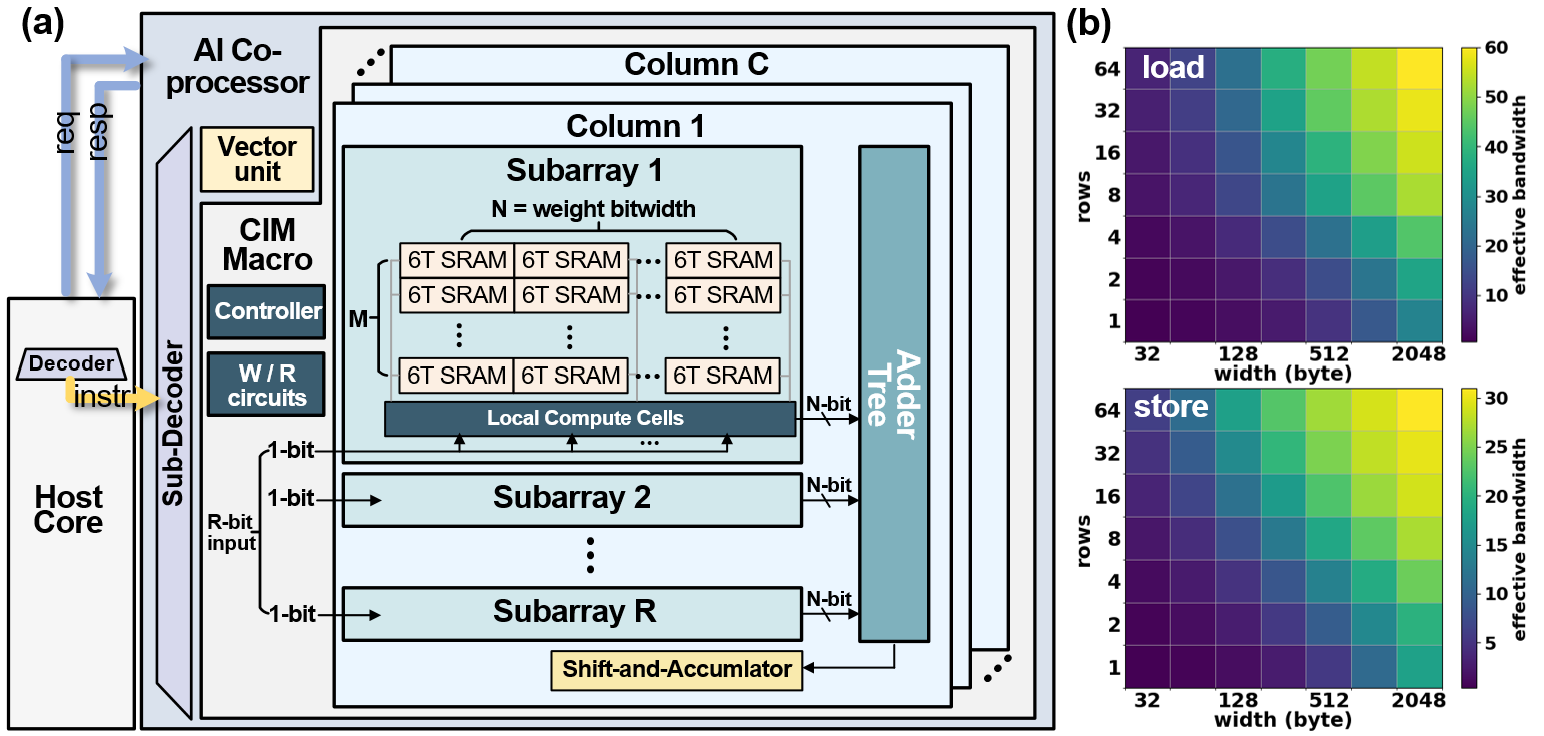}
  \vspace{-5 pt}
  \caption{Mem-centric core with CIM macro design.}
  \label{fig:cim}
  \vspace{-5 pt}
\end{figure}

%\begin{figure}[t]
%  \centering
%  \includegraphics[width=0.95\linewidth]{fig5_1.png}
%  \caption{Comparison of Solution Paths for AI Acceleration}
%  \label{fig:Heterogeneous Cores}
%  \vspace{-15 pt}
%\end{figure}
Fig.~\ref{fig:systolic} and Fig.~\ref{fig:cim} illustrate the design details of CC and MC cores. Both core types consist of a host core and a coprocessor. The extended instructions are decoded by host core and dispatched to coprocessor via direct-linked interface. %The CC-core employ a systolic array, which is wide used  as accelerator for GEMM calculation, and memory-central core utilize CIM macro being suitable for GEMV. 

\textbf{Compute-Centric core:} As shown in Fig.~\ref{fig:systolic}, the bulk of compute-centric GEMM coprocessor is a systolic array, which contains \textit{R}\(\times\)\textit{C} %BF16-INT8 
multiply-accumulate processing elements (PE). For GEMM operations, the weights stored in the PE array remain stationary, while the input activations are sent into the array in a systolic manner. %However, in the case of GEMV, there is only one column of activation (circled in red in fig~\ref{fig:systolic}), which can cause a serious drop in PE utilization. This is one motivation to adopt heterogeneous architecture to avoid relying on systolic arrays for GEMV. 
Four \textit{R}\(\times\)\textit{C} matrix registers are equipped to store operands. %As the SA lacks the capability for element-wise computation, 
The vector units are employed to execute vector instructions for element-wise computations. Vector instructions share the matrix registers and have an element width of \textit{C}, enabling parallel operation on a row of a matrix register by one instruction.
%Matrix and vector instructions share the same register file. The element width of vector units equals \textit{C}, enabling the parallel operations on a row of matrix registers in a single instruction.%The vector unit shares the matrix registers %with systolic array for data transmit. %to calculate addition, multiplication or shift. 
%For one instruction, the vector unit can process the operation for one row of a matrix register, which consists of \textit{C} INT8 elements. 
The coprocessor has an independent load/store unit for matrix registers, avoiding the data bottleneck caused by the narrow port of the host core. 
For multiplication between a \textit{R}\(\times\)\textit{C} and an \textit{M}\(\times\)\textit{R} matrix, the cycles of loading and computing equals: 
\begin{equation}
\label{LSA}
L_{SA}=R+(R-1)+(C+M-1)-1=2R+C+M-3
\end{equation}
Systolic dataflow enables data transfer solely between neighboring PEs, reducing wiring overhead. 
However, there is only one column of activation (circled in red in Fig.~\ref{fig:systolic}) in GEMV, causing PEs idleness and inefficiency.   

\textbf{Memory-Centric core:} As shown in Fig.~\ref{fig:cim}, the main feature of an MC-core is utilizing digital CIM for the coprocessor, which integrates compute cells within SRAM macro to avoid reg-to-SRAM load/store. %Based on digital CIM technology, the compute units are merged into data SRAM to form a CIM macro. 
A CIM macro consists of a controller, wordline decoder, write-and-read circuits, and \textit{C} columns. Each column contains \textit{R} subarrays, an adder tree, and a shift-and-accumulator. Each subarray consists of \textit{M}\(\times\)\textit{N} 6T SRAM bit-cells. Here \textit{N} equals to the bit-width of a weight data. During GEMV, %one of the \textit{M} weights in a subarray is read into compute cells as the multiplier. 
the \textit{W}-bit activation data is broadcasted into the columns in bit-serial. %and multiplied with one of \textit{M} weight in subarrays. %In our design we let \(N=8\), \(W=8\) to perform INT8\(\times\)INT8. 
One of the weights in each subarray is read out and multiplied with 1-bit activation every cycle. 
Adder tree sums the \textit{R} product and sends the result to shift-and-accumulator, in which the sum is shifted and accumulated to finish the entire \textit{W}-bit multiplication. The broadcast dataflow fully utilizes the compute cells for GEMV, which is completed in \textit{W+1} cycles.
%and won't have excessive wiring overhead for bit-serial input. 
The cycle number of a \textit{M} rows GEMM is:
\begin{equation}
\label{LCIM}
L_{CIM}=M*W+1
\end{equation}
Compared to SA, CIM requires fewer cycles for GEMV, i.e. $M=1$ in (\ref{LSA})(\ref{LCIM}). But for GEMM, CIM is less efficient due to the bit-width factor \textit{W}.
%For GEMM, CIM  \(M\) is not a small number, CIM macro takes more cycles because of factor (\(W\) )
Moreover, MC-clusters have significantly larger data memory than CC-clusters, which allows larger matrix block to be transferred at once. 
%cannot increase the bandwidth but 
%can enhance memory access patterns. %enhance the effective bandwidth for data transfer. 
%Larger memory allows larger 2D matrix block to be transferred at once, 
%The extended continuous data access enhances DMA and DRAM efficiency. 
We assess the effective bandwidth (data amount/cycles) for transferring matrices with various sizes. 
Fig.~\ref{fig:cim} (b) shows that the effective bandwidth drops notably for small matrices, but nears the ideal bandwidth as matrix size increases. This indicates the ample on-chip memory in MC-cluster can alleviate the bandwidth pressure and enhance DMA and DRAM efficiency. %for memory-bound GEMV.

%We further evaluate the latency of GEMM and GEMV with typical size in real AI task for one CC-cluster and one MC-cluster. The DRAM-to-chip data transfer is also considered. The performance of CC-cluster is 4.22\(\times\) better than MC-cluster for GEMM. For GEMV, the MC-cluster is 1.87\(\times\) faster than CC-cluster.

\subsection{ISA Extension and Programming Model}
\begin{figure}[t]
  \centering
  \includegraphics[width=\linewidth]{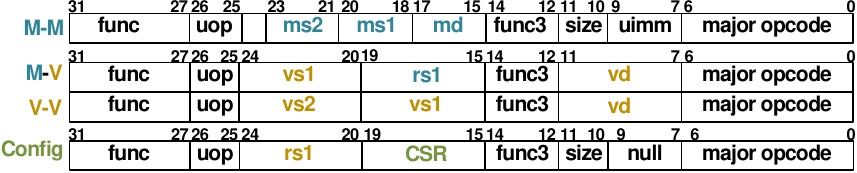}
  \vspace{-15pt}
  \caption{Instruction format of the extended instructions.}
  \label{fig:Instr}
  \vspace{-10 pt}
\end{figure}

We adopt RISC-V as our main ISA and extend instructions for AI acceleration as shown in Fig~\ref{fig:Instr}. 
The first row illustrates the matrix instruction (i.e. mm\_mul for GEMM) supported by CC-core. 
The matrix instructions differ from scalar instructions by employing matrix registers for both the source and destination operands. 
The CIM-based MC-core supports matrix-vector instructions, in which the \textit{vs1} and \textit{vd} are vector registers that store source and destination operands, and the \textit{rs1} store the base address of matrix operand. The vector units in all cores support a subset of RISC-V vector instructions for activation functions and data precision conversion. Configure instructions can modify the control and status registers (CSR), storing the runtime parameters such as vector/matrix sizes.

The programming model of our EdgeMM is similar to snich\_cluster \cite{zaruba2020snitch}. 
%On the memory level, given that most data objects in AI workloads are large 2D matrices, complex multi-level data caches are avoided. 
%As mentioned in Section 3, software-managed shared data memory and DMA module are an alternative. 
The computing tasks, e.g. GEMM/GEMV, are allocated across cores with tensor partitioning. Specifically, each core and cluster contains read-only CSRs that store its index and type, allowing cores to calculate address offsets to access the allocated tensor shards. The core synchronization function is also supported. %allowing software to allocate computations across cores directly. 
For ISA extension, the extended instructions can be utilized by customized kernel functions, enabling the use of the RISC-V toolchain without the internal modification of the compiler.

\section{Bandwidth Optimizations }
%Unlike computational resources, bandwidth resources are more strictly constrained and shared among all cores. 
To effectively utilize the limited bandwidth, we optimize memory access reduction by activation-aware weight pruning and perform token-length-driven bandwidth management.

\subsection{Activation-aware Weight Pruning for GEMV}
\begin{figure}[t]
  \centering
  \includegraphics[width=\linewidth]{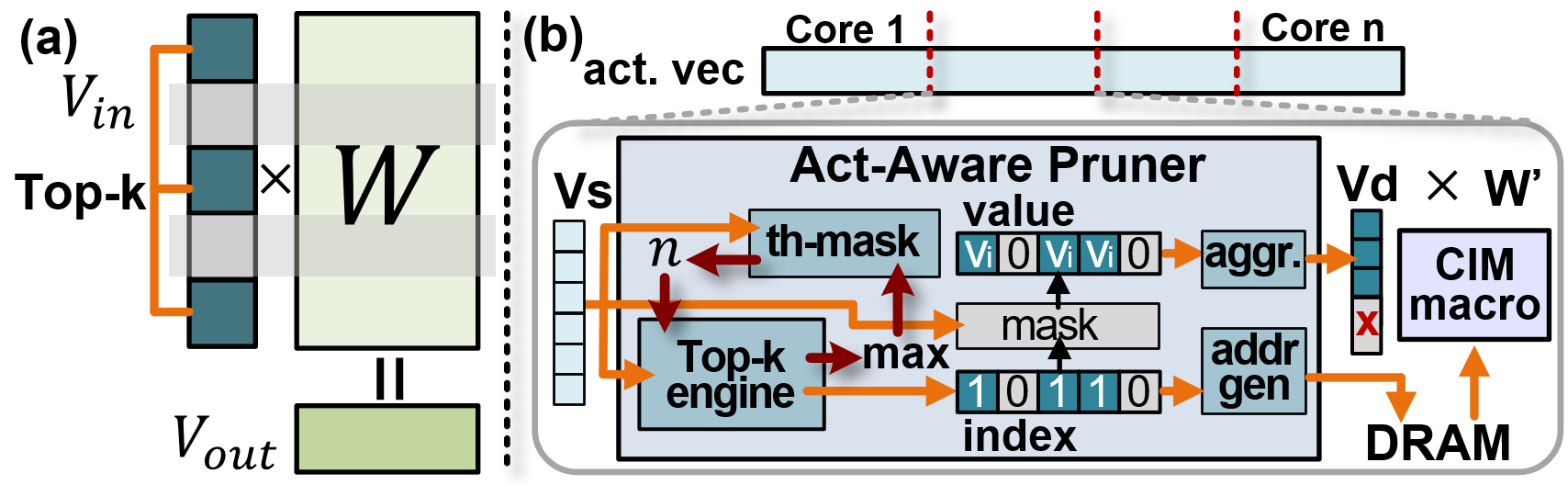}
  \vspace{-15 pt}
  \caption{Activation-aware weight pruning and hardware pruner.}
  \label{fig:act_pruning}
  \vspace{-5 pt}
\end{figure}

%To reduce data transfer from DRAM and accelerate the LLM-decoding, an activation-aware weight pruning is introduced.  
%The values in the weight matrix are relatively even, while the activation vector contains some outliers and sparse channels, as shown in fig~\ref{fig:activation}. Thus, 
As Sec.~\ref{sec:profiling} mentioned, the FFN weight matrices dominate DRAM access, and the activation vectors exhibit significant sparsity and outliers across channels. 
This motivates us to perform channel-wise activation-aware pruning. 
As shown in Fig.~\ref{fig:act_pruning}, GEMV can be performed by selecting the Top-\textit{k} outliers and pruning the rest channels with corresponding rows in weight matrices.
Recent works \cite{wanda,lee2024cats} adopt the similar concept, but they all regard \textit{k} as a fixed empirical input.
%, which is essential for balancing optimization effect and result accuracy, . 
In fact, \textit{k} can vary along different decoder layers for better pruning.
%in a token generation of LLM. 
For example, early decoder layers should retain most weights to avoid errors accumulating and the latter layers have more distinct outliers, allowing for more efficient pruning. 
%in following layers. As layer output prediction is more definitive, with outliers are much more distinct, 

%GEMV is dominated by outliers, enabling computation by selecting only the Top-\textit{k} channels of the vector, thereby pruning the sparse channels and corresponding rows of the weight matrix 

%during token generation, as the layer depth increases , the selection output token becomes more definitive, with outliers standing out more distinctly. 
%In this case, more channels can be pruned by decreasing \textit{k}. %Conversely, in the shallower layers k should remain large to maintain accuracy.
%Based on this sparsity, GEMV can be performed by selecting only the top-\textit{k} channels of the vector, pruning the sparse channels and corresponding rows of the weight matrix. 
%Recent works adopt the same idea, but they all regard \textit{k}, which is essential for balancing optimization effect and result accuracy, as a fixed empirical input. 
%The \textit{k} should adapt dynamically based on the data distribution of the vector. 

As shown in Alg.~\ref{algo_topk}, a layer-wise dynamic Top-\textit{k} pruning scheme is presented. %As shown in Fig~\ref{fig:scheduling} (a.1), it is unnecessary to read and multiply the matrix rows corresponding to the sparse channels.
%thus reducing both computation and memory access. 
The update of \textit{k} is based on straightforward yet effective principles: a channel that is \textit{t} times smaller than the max channel can be considered negligible, where \textit{t} is a fixed parameter set to 16 in our design. Moreover, \textit{k} should decrease progressively with layer depth. If the number of retained channels \textit{n} is less than \textit{k}, it indicates that the current \textit{k} is too large. Thus, \textit{k} is set to \textit{n}.%If the number of negligible channels is less than the current k, then update k to n to let k decrease with the layer depth, as shown in Algorithm 1.
%\afterpage{
%The \textit{t} is set to 16 in our design. The Algorithm 1 can be applied to all GEMV in FFN. Selection of Top-\textit{k} and max channels on high dimension vectors is complex.

\IncMargin{1em}
\begin{algorithm}
  \SetKwData{Left}{left}\SetKwData{This}{this}\SetKwData{Up}{up}
  \SetKwFunction{Union}{Union}\SetKwFunction{FindCompress}{FindCompress}
  \SetKwInOut{Input}{input}\SetKwInOut{Output}{output}
  \Input{Threshold $t$, Vector dim $d$}
  %\Output{A partition of the bitmap}
  %\BlankLine
  %$k = k_{0}$   \tcp*[r]{init the k}
  \For{ \textit{layers} in model}{
  %/* previous computation */\\
    %Self-Atten() \tcp*[r]{prior irrelevant computation}
    %/*    FFN:    */ \\
    \If{$layer\_index$ == 1}{
        $k$ = $d$\tcp*[r]{no pruning}
        %continue
    }
    $V_{x}^{'}$ = top-k($V_{x}$, $k$) \tcp*[r]{select channels}\
    $W^{'}$ = pruning($W$, index($V_{x}^{'} $))\tcp*[r]{pruning}\
    GEMV($W^{'}$, $V_{x}^{'}$)\\
   % \tcp{num of non-negligible channels:}\
    $n$ = count($V_{x}[i]$ $>$ $\frac{max(V_{x}[i] )}{t}$ )\\ 
    \If{$n < k$}{
        $k$ = $n$ \tcp*[r]{update $k$}
    }
    /* remaining computation */\\
  }
  generate an output token \\
  %$k$ = $k_{0}$ \tcp*[r]{reset the $k$}
  \caption{Layer-wise Dynamic Top-k Pruning}\label{algo_topk}
\end{algorithm}%\DecMargin{1em}

The hardware pruner in the MC-core facilitates the execution of GEMV with Alg.~\ref{algo_topk}, as shown in Fig~\ref{fig:act_pruning} (b). In practice, the activation vector is allocated to cores by channels. Each core focuses on its assigned local channels, avoiding complex global Top-\textit{k} selections. 
%Since data is evenly distributed across channels, localization has little impact on pruning effectiveness. 
The core first loads a slice of the activation vector into the vector register \textit{vs}, then calls the hardware Act-Aware pruner for processing via specific instruction. In pruner, the Top-\textit{k} engine finds the \textit{max} value in \textit{vs} and marks the selected elements with 1 in the index register. The th-mask receives the \textit{max} value and \textit{vs}, generating \textit{n} in Alg.~\ref{algo_topk} for \textit{k} update. The index is fed into the address generator, which calculates the address of the non-pruned weight rows for DRAM read requests. Following that, the \textit{vs} is masked by the index and aggregated, and then output to \textit{vd} register. Then the CIM macro computes the GEMV between \textit{vd} and stored weights. The above process reduces DRAM access without wasting on-chip computation or memory resources. 

\subsection{Token-Length-Driven Bandwidth Management}

The heterogeneity of EdgeMM complicates workload distribution. For MLLMs, it is optimal to run modality encoder and LLM-prefill on CC-clusters, with LLM-decoding on MC-clusters. In real-time applications (e.g. AD, robot, AR/VR), continuous streaming input enables pipelined execution. But the latency of LLM-decoding increases with the output token length, which can result in an unbalanced pipeline. %as shown in Fig~\ref{fig:scheduling}(a).

\begin{figure}[t]
  \centering
  \includegraphics[width=\linewidth]{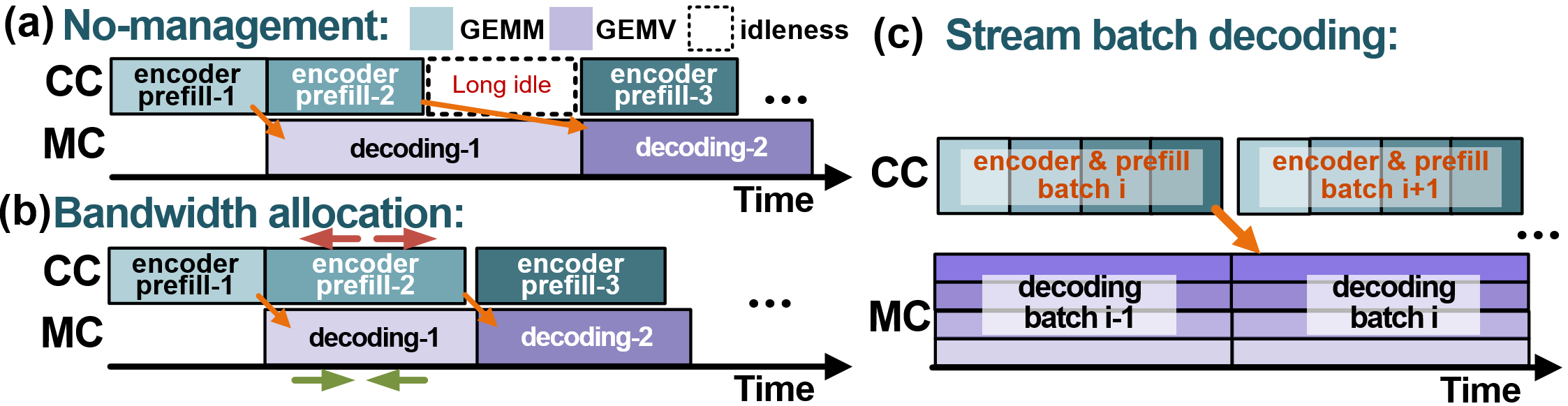}
  \vspace{-20 pt}
  \caption{Pipeline of bandwidth and workload management.}
  \label{fig:management}
  \vspace{-10 pt}
\end{figure}

To maintain a balanced pipeline for various token lengths \textit{l}, a throttling-based dynamic bandwidth allocation is adopted. 
%\cite{memguard,E-warp}
%\textit{l} can be predicted by models or estimated from runtime history. 
For a model with default equal bandwidth sharing among clusters, an expected token length $l_{e}$ can balance the latency of CC and MC clusters. 
As shown in Fig.~\ref{fig:management}(a), if $l$ exceeds $l_{e}$, CC-cluster will enter waiting idle. 
In such scenarios in Fig.~\ref{fig:management}(b), limiting CC-cluster bandwidth and reallocating it to MC-clusters can accelerate LLM-decoding, reducing overall latency. 
%While the bandwidth allocation slows down the MC clusterselapses and PMCs are reset, the overall latency is still shortened because MC clusters are not the bottleneck in pipeline. 
%To implement bandwidth allocation, 
To implement bandwidth allocation, each cluster is assigned a memory access budget \textit{B}. Within an interval \textit{T}, the performance monitoring counter (PMC) in DMA accumulates the memory access usage \textit{d} of each cluster. If $\textit{d}>\textit{B}$, subsequent DMA requests from this cluster are blocked until \textit{T} elapses and PMCs are reset. We adjust the cluster budget $B_{c}$ and $B_{m}$ to enable dynamic bandwidth allocation.

%By default, all clusters have equal priority for off-chip bandwidth sharing,

%$B_{cc}$ should be reduced to limit memory access, allocating more bandwidth resources to MC-clusters as shown in fig (c).%The allocation operates on a interval of T. 
%To enable varied bandwidth allocation,  %B for CC-clusters and MC-clusters can be adjusted for dynamic bandwidth allocation.%based on the output token length, enabling dynamic bandwidth allocation.

%Off-chip bandwidth is shared by all clusters, and increasing the bandwidth allocated to a cluster can accelerate its processing, particularly for memory-bound GEMV.

%The proposed dynamic bandwidth allocation relies on output token lengths, which can be predicted by models or estimated from runtime history.  
Note the bandwidth allocation can only balance the pipeline with \textit{l} in a certain range. For larger \textit{l}, stream-based batch processing can be employed. As shown in Fig~\ref{fig:management} (c), CC-clusters can encode and prefill a batch of streaming input without waiting for LLM-decoding. The MC cluster performs decoding concurrently on this batch. 
Larger batch multiplies the latency of CC-clusters, while decoding latency of MC-clusters increases slightly for the weight reuse. Hence the overall throughput can be boosted with minimal latency loss.

 %the bandwidth occupation of CC-clusters should be restricted, and more bandwidth resources should be allocated to MC-clusters.

\section{Evaluations}

\begin{figure}[t]
  \centering
  \includegraphics[width=\linewidth]{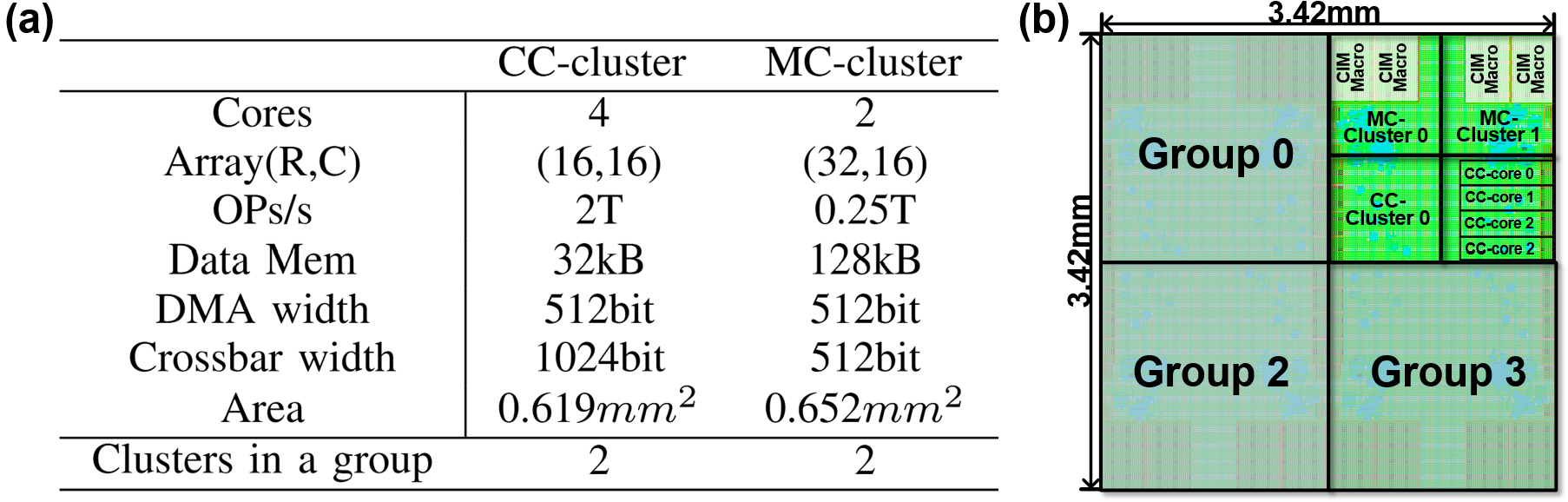}
  \vspace{-15 pt}
  \caption{Design configurations and the 22nm layout of EdgeMM.}
  \label{fig:layout}
  \vspace{-10 pt}
\end{figure}

%\subsection{Evaluation Method}
\subsection{Implementation and Evaluation Setup}
We leverage open-source Snitch\_cluster\cite{zaruba2020snitch} as the foundation platform for our EdgeMM development. 
The CC-cluster and MC-cluster are implemented using RTL flow and in-house 22nm digital CIM macro IP.
%The configuration and comparison of CC-cluster and MC-cluster are shown in Table ~\ref{tab:cluster_table}.
We utilize Cadence Genus and Cadence Innovus for synthesis and physical implementation at 1GHz using TSMC 22nm technology.
Fig.~\ref{fig:layout} demonstrates the layout view and design configurations of EdgeMM. The chip power is 112mW according to post-P\&R report. The SA-based coprocessor occupies 62\% area of a CC-core, and the CIM macro occupies 81\% area of an MC-core. 
The performance is evaluated by RTL simulation and the in-house simulator with a dedicated mapping explorer. %and the SPHINX-Tiny \cite{gao2024sphinx} is the target MLLM.

%To evaluate the performance, we combine RTL simulation with the in-house modeling tool and a dedicated mapping explorer. 

%A CC-cluster and an MC-cluster have comparable total areas, yet their resource distribution differs greatly. In CC-clusters, the SA-based coprocessor occupies 62\% area of a core, with 4 cores sharing 32kB of data memory, demonstrating a high computational density. In MC-clusters, the CIM macro within the coprocessors occupies the majority of the area, providing 128kB of data memory.

 %The designs are synthesized by Cadence Genus and physically implemented by Cadence Innovus at 500MHz using TSMC 22nm technology.

% \begin{table}[t]
%       \caption{Cluster Configuration}
%     \label{tab:cluster_table}
%     \vspace{-10 pt}
%   \resizebox{\linewidth}{!} {
% \begin{tabular}{lccllc}
% \hline
%                     & \multicolumn{1}{l}{Cores(n)} & \multicolumn{1}{l}{Data Memory} & array(R,C) & TOP/s                  & \multicolumn{1}{l}{Crossar bw} \\ \hline
% \textbf{CC-cluster} & 4                            & 24kB                            & (32.32)    & \multicolumn{1}{c}{2T} & 1024bit                        \\
% \textbf{MC-cluster} & 2                            & 128kB                           & (32,32)    & 0.128T                 & 512bit                         \\ \hline
% \end{tabular}
% }
% \vspace{-10pt}
% \end{table}

\subsection{Benefits of Heterogeneous Extensions}
\begin{figure}[t]
  \centering
  \includegraphics[width=\linewidth]{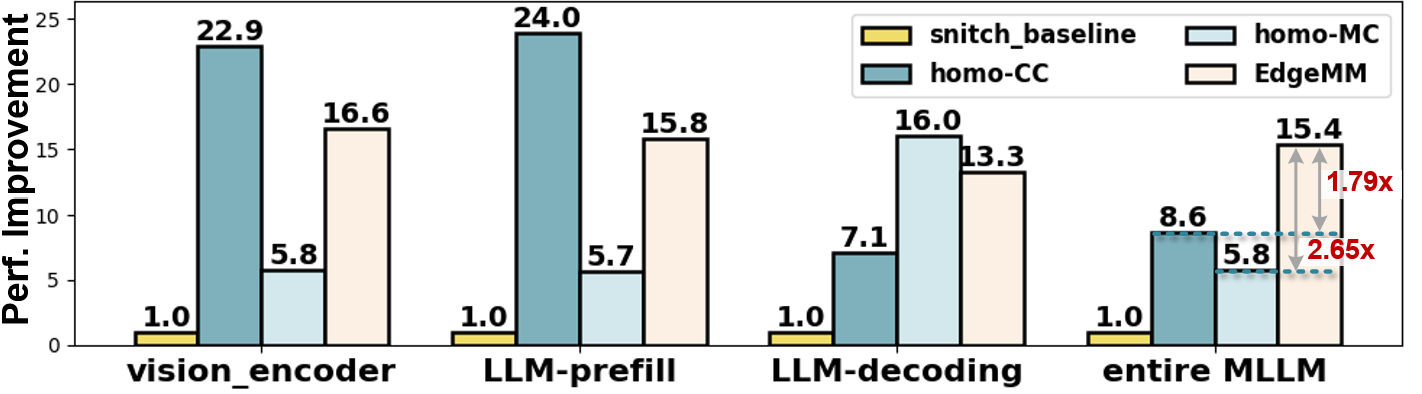}
  \vspace{-15 pt}
  \caption{Perf. improvements of homo. and hetero. designs}
  \label{fig:hete}
  \vspace{-5 pt}
\end{figure}

To evaluate the benefits of our heterogeneous AI-extensions, we compare our heterogeneous EdgeMM with homogeneous cluster designs: homo-CC and homo-MC, which consist solely of CC-clusters and MC-clusters. 
The original snitch\_cluster \cite{zaruba2020snitch} including SIMD cores is set as the baseline. 
%We also compare the performance with multi-core solution with homogeneous extension, 
%We compare the performance of the above hardware designs for 
The SPHINX-Tiny \cite{gao2024sphinx} MLLM and its inner phases with averaged input and output token length are used as the workload.

As shown in Fig.~\ref{fig:hete}, all extended designs have significant performance boosts compared to the baseline since matrix extensions and coprocessors enhance computing power and reduce redundant load/store.
%compared to conventional SIMD cores. 
Homo-CC and homo-MC perform the peak speedup in their adapted phases: a CC-cluster shows 4.3$\times$ better GEMM performance than an MC-cluster, whereas an MC-cluster is 2.42$\times$ faster in GEMV due to more efficient memory access.
However, homo-CC and homo-MC encounter bottlenecks in non-adapted phases due to hardware mismatch. For the entire MLLM, the heterogeneous EdgeMM outperforms them by 1.79$\times$ and 2.65$\times$ respectively. %demonstrating the benefits of heterogeneity.

%The evaluation reveals that two types of clusters exhibit significant performance differences, and heterogeneity provides substantial benefits. Weight pruning is excluded from the subsequent experiments.  
%For LLM-prefill and vision\_encoder which share similar model structures and attributes, the homogeneous design composed entirely of CC-clusters achieves the highest speedup ratio due to its massive computational resources. 

\subsection{Benefits of Bandwidth Optimizations}
\begin{figure}[t]
  \centering
  \includegraphics[width=\linewidth]{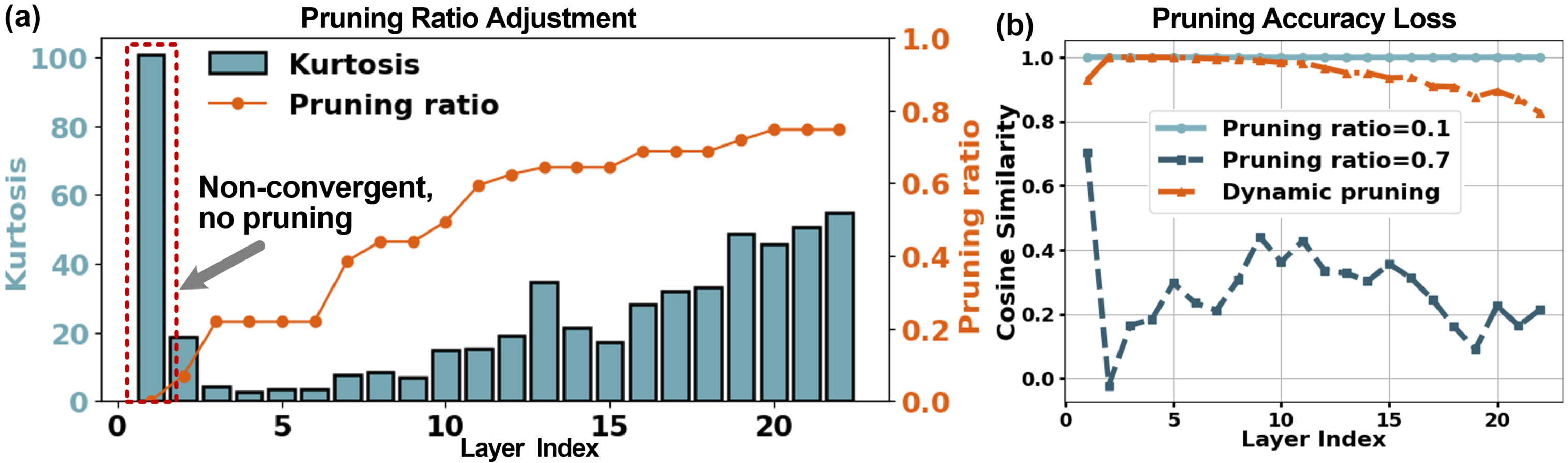}
  \vspace{-15 pt}
  \caption{Evaluation of activation-aware dynamic Top-k weight pruning.}
  \label{fig:pruning_eva}
  \vspace{-15 pt}
\end{figure}

%%%%%%%%%%%%%%%%%%%%%%%%%%% Changed by Mingxuan Li %%%%%%%%%%%%%%%%%%%%%%%%%%%
%%% 从直觉上来说，pruning rate 指的是被剪枝的数的百分比，即 pruning rate 越大，数据越稀疏
%%% 图中的 pruning rate 似乎和直觉相反
%%% 我建议按照直觉上的 pruning rate，这样也不用解释 pruning rate 是什么了
%%% 能否第一层的 pruning rate 在图上就设置成没有剪枝？说明我们注意到了第一层的重要性，将其排除
%%% 感觉（b）缺少 dynamic pruning 策略的具体剪枝比例的数据？
%%% 我建议画两个 fixed pruning 曲线，其中一条精度和我们的动态方法接近，另一条剪枝比例和我们的方法接近，这样能说明：我们在使用了平均剪枝比例较大的方法后，精度可以和剪枝比例较小的fix prune 方法媲美
Fig. \ref{fig:pruning_eva} (a) presents the evaluation results for activation-aware dynamic TOP-\textit{k} weight pruning during a token generation, where \textit{Kurtosis} is a mathematical metric to represent channel-wise data distribution.
Higher \textit{Kurtosis} indicates more distinct outliers, i.e. more channels can be pruned.
Results show that with \textit{Kurtosis} increasing with layer depth, with pruning ratio of a randomly chosen core increases dynamically, proving the activation-awareness of dynamic TOP-\textit{k} mechanism.
Note that although the first layer exhibits a high \textit{Kurtosis}, its data distribution remains unstable. Our experiment reveals that pruning this layer leads to significant accuracy degradation.
Thus Alg.~\ref{algo_topk} skips the first layer.

To validate the accuracy of dynamic TOP-\textit{k} pruning, we assess the cosine similarity between pruned and unpruned output vectors in FFN as depicted in Fig.\ref{fig:pruning_eva} (b).
It is observed that dynamic pruning achieves comparable accuracy as a mild fixed pruning ratio of 0.1 while fixing pruning ratio at 0.7 causes irreversible accuracy loss in the shallow layers. 
Overall, our activation-aware pruning reduces the LLM-decoding latency of SPHINX-Tiny by 42\% on average, with minimal score reduction in VQA \cite{VQA}.

\begin{figure}[t]
  \centering
  \includegraphics[width=\linewidth]{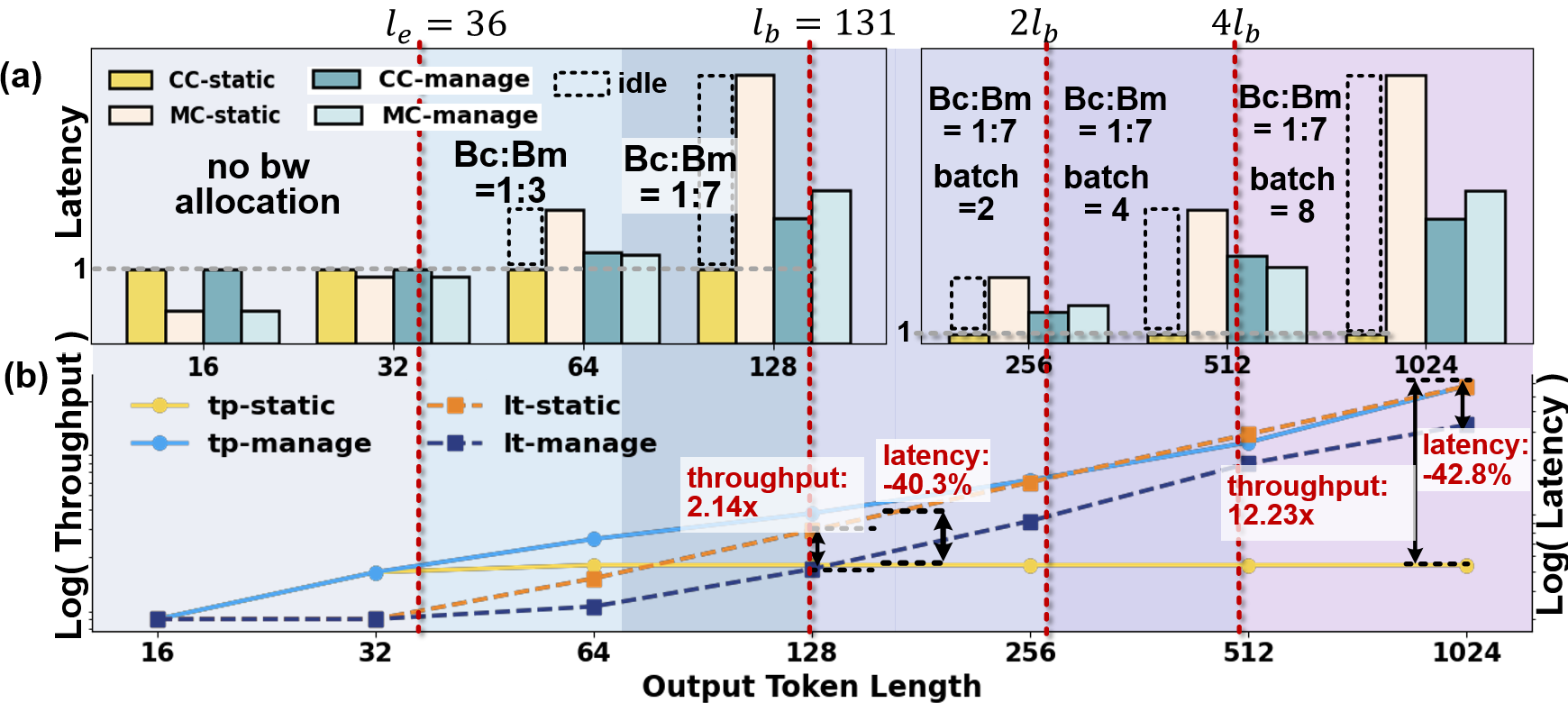}
  \vspace{-10 pt}
  \caption{(a) Latency and (b) throughput gains by bandwidth management.}
  \label{fig:bw_opt}
  \vspace{-10 pt}
\end{figure}

%%%%%%%%%%%%%%%%%%%%%%%%%%% Changed by Mingxuan Li %%%%%%%%%%%%%%%%%%%%%%%%%%%
%%% 感觉没必要分为（a）（b），文字都是在一起讲，看图也是一起从左向右看
Our dynamic bandwidth allocation can harness the heterogeneous cores under varying output token lengths \textit{l}.
As described in Fig. \ref{fig:bw_opt}, when $l<l_{e}=36$, bandwidth is not the critical bottleneck as the latency of CC-cores is large.
As \textit{l} rises, the extended duration of LLM-decoding slows down the pipeline.
Thus the bandwidth budget of CC-cores $B_{c}$ is reduced progressively, reallocating bandwidth to MC-cores.
The $B_{c}$:$B_{m}$ ratio ranges to 1:3 or even 1:7, accelerating MC-cores with tolerable latency increment of CC-cores.
This optimization reduces latency by 40.3\% and improves throughput 2.14$\times$ at \textit{l}=128.
%capable of covering most edge scene.
When \textit{l} increases to $l_{b}=131$, bandwidth allocation nears its limit, where batch decoding is introduced to balance the pipeline.
While batch decoding increases latency by 42\% at $l=1024$, it boosts throughput by 13.98$\times$, compensating for the latency overhead.
We further compare EdgeMM with existing mobile GPU, as shown in Table \ref{tab:gpu}. 
%under same bandwidth and moderately higher computational power, achieves significant throughput improvements compared to typical. 
Although SM cores in GPU provide massive parallelism, they often remain underutilized and not as effective as SA and CIM extensions in EdgeMM. 
%Moreover, tensor cores in GPU often remain underutilized during the LLM-decoding, alongside bandwidth is not fully used in encoder and prefill phases of MLLMs. 
%With similar resources, heterogeneous design and effective management enhance resource utilization and boost throughput. 
Compared to laptop RTX 3060 GPU, EdgeMM achieves 2.15$\times$ performance for better resource utilization driven by heterogeneity. With our activation-aware weight pruning, the performance gain can be enhanced to 2.84$\times$, reaching 138 tokens/s throughput. Overall, EdgeMM achieves a 0.28 token/J energy efficiency. %\textcolor{red}{which is xx orders of magnitude higher than GPU.}

\vspace{-10pt}
\begin{table}[h]
       \caption{Comparison of EdgeMM and mobile GPU}
     \label{tab:gpu}
     \vspace{-5 pt}
   \resizebox{\linewidth}{!} {
\begin{tabular}{lcccl}
\cline{1-4}
                                    & \multicolumn{1}{l}{Compute Power} & \multicolumn{1}{l}{Bandwidth}                                            & \multicolumn{1}{l}{MLLM perf.} &  \\ \cline{1-4}
\multicolumn{1}{c}{RTX 3060 Laptop} & 13TFLOP/s (FP32)                        & \multirow{3}{*}{\begin{tabular}[c]{@{}c@{}}GDDR6\\ 336GB/s\end{tabular}} & 1$\times$                             &  \\
\multicolumn{1}{c}{EdgeMM}          & 18TFLOP/s (BF16)                        &                                                                          & 2.15$\times$                           &  \\
EdgeMM+Weight pruning               & 18TFLOP/s (BF16)                         &                                                                          & 2.84$\times$                           &  \\ \cline{1-4}
\end{tabular}
}
\vspace{-5pt}
\end{table}

% \begin{table}[t]
%        %\caption{Cluster Configuration}
%      \label{tab:cluster_table}
%      \vspace{-10 pt}
%    \resizebox{\linewidth}{!} {
% \begin{tabular}{ccc}
% \hline
% \multicolumn{1}{l}{}                                          & \multicolumn{1}{l}{CC-cluster} & \multicolumn{1}{l}{MC-cluster} \\ \hline
% \multicolumn{1}{c|}{Cores}                                    & 4                              & 2                              \\
% \multicolumn{1}{c|}{Array(R,C)}                               & (16,16)                        & (32,16)                        \\
% \multicolumn{1}{c|}{OPs/s}                                     & 2T                             & 0.25T                           \\
% \multicolumn{1}{c|}{Data Mem}                                 & 32kB                           & 128kB                          \\
% \multicolumn{1}{c|}{DMA width}                                & 512bit                         & 512bit                         \\
% \multicolumn{1}{c|}{Crossbar width}                           & 1024bit                        & 512bit                         \\
% \multicolumn{1}{c|}{Area}                                     & 0.619$mm^{2}$                        & 0.652$mm^{2}$                          \\ \hline
% \begin{tabular}[c]{@{}c@{}}Clusters in a group\end{tabular} & 2                              & 2                              \\ \hline
% \end{tabular}
% }
% \vspace{-10pt}
% \end{table}

\section{Conclusion}
We present EdgeMM, a multi-core CPU with heterogeneous AI-extension for edge MLLM. To accommodate phases with diverse attributes in MLLM, coprocessors based on systolic arrays and digital CIM macros are integrated for GEMM and GEMV. To leverage limited bandwidth, activation-aware dynamic weight pruning is proposed with hardware support. Token-length-driven bandwidth allocation is employed to fully utilize heterogeneous cores. Overall, EdgeMM achieves 2.84$\times$ performance improvement compared to RTX 3060 GPU.

%about 138 token/s and 0.28 token/J, which is

%\section*{Acknowledgment}
%The preferred spelling of the word ``acknowledgment'' in America is without an ``e'' after the ``g''. Avoid the stilted expression ``one of us (R. B. G.) thanks $\ldots$''. Instead, try ``R. B. G. thanks$\ldots$''. Put sponsor acknowledgments in the unnumbered footnote on the first page.

%\clearpage

\section*{Acknowledgment}

This work is supported in part by NSFC Grant No. 92464202 and CIE-Smartchip research fund.

\bibliographystyle{ieeetr}
\bibliography{reference}

\end{document}